\documentclass[11pt,a4paper]{scrartcl}

\usepackage[utf8]{inputenc}
\usepackage[english]{babel}           
\usepackage{amsfonts, amsmath, amsthm, amssymb, dsfont}
\usepackage{graphicx}
\usepackage{epstopdf}
\usepackage{color}
\usepackage{hyperref}

\usepackage{lineno}

\usepackage[numbers]{natbib}

\newcommand{\s}{\footnotesize}  %controll the font size of figure labels
\title{On strongly nonlinear gravity waves in a vertically sheared atmosphere}
\subtitle{Part I: Spectral stability of the refracted wave}
\author{Mark Schlutow\footnote{mark.schlutow@fu-berlin.de}\\ 
\s Institut f\"ur Mathematik, Freie Universit\"at Berlin, Germany\\
and\\ Georg S. V\"olker\\
\s Institut f\"ur Atmosphäre und Umwelt, Goethe Universität Frankfurt am Main, Germany}
\date{}

\begin{document}

	\maketitle

	\begin{abstract}
		We investigate strongly nonlinear stationary gravity waves which experience refraction due to a thin vertical shear layer of horizontal background wind.
The velocity amplitude of the waves is of the same order of magnitude as the background flow
and hence the self-induced mean flow alters the modulation properties to leading order.
In this theoretical study, we show that the stability of such a refracted wave depends on the classical modulation stability criterion
for each individual layer, above and below the shearing.
Additionally, the stability is conditioned by novel instability criteria providing bounds on the mean-flow horizontal wind and the amplitude of the wave.
A necessary condition for instability is that the mean-flow horizontal wind in the upper layer is stronger than the wind in the lower layer.
	\end{abstract}

	\section{Introduction}

The importance of gravity waves for the atmospheric dynamics 
and hence for weather and climate forecasting was first established in \cite{Dunkerton1978} and \cite{Lindzen1981}. 
Gravity waves redistribute energy vertically 
and thereby couple the different layers of the atmosphere \citep{Fritts2003,Becker2012}. 
Usually excited in the troposphere, gravity waves may persist deep into the upper atmospheric layers \citep{Fritts2016,Fritts2018,Fritts2019}
where they interact with the mean flow. They exert drag onto the horizontal mean-flow momentum, produce heat when dissipating
\citep{Becker2004}, and cause increased mixing of tracer constituents such as green-house gases \citep{Schlutow2014}.
To this day, many questions regarding the sources, propagation and dissipation of gravity waves are still to be answered.

From a conceptional point of view, three distinct types of theories---each with its own benefits and challenges---may be employed to study these questions.
First and foremost, linear wave theory provides the foundation for our understanding of gravity waves. It predicts dispersion, refraction, reflection, wave packet formation etc. as long as the amplitudes of the waves can be considered to be infinitesimally small \cite{Bretherton1966}. 

For large-amplitude gravity waves, perturbation theory is a versatile and powerful tool leading to weakly nonlinear wave theory. 
The key idea of weakly nonlinear theory is to expand the linear solution to the fluid dynamical equations 
in an asymptotic series assuming small but finite amplitude. 
Corrections of the order of the amplitude squared to the linear model add new phenomena to wave theory such as wave-mean-flow interaction and modulational instabilities \cite{Whitham1974a,Grimshaw1977,Sutherland2001,Sutherland2006,Tabaei2007a}.
For internal gravity waves in the oceans, weakly nonlinear theory is almost always sufficient. % as amplitudes are typically much smaller than the background flow.
However, this is often not the case for atmospheric gravity waves.
When excited over mountains, tropospheric gravity waves may have amplitudes of the order of magnitude of the background wind. 
In the stratosphere and higher, such strong amplitudes are rather the rule than the exception.
This effect is due to anelastic amplification which is absent in the ocean and caused by the compressibility of air.
The majority of atmospheric gravity waves is excited in the troposphere. 
When they extend into higher altitudes, they encounter exponentially decreasing background density.
It is then energy conservation that lets the waves grow exponentially causing strong amplitudes.

The mathematical description of gravity waves beyond weakly nonlinear theory was pioneered by Grimshaw \cite{Grimshaw1972,Grimshaw1974} 
and further analyzed in \cite{Achatz2010} and also \cite{Schlutow2017b} in terms of modulation equations.
In this asymptotic theory the small expansion parameter is the ratio between the scale 
on which the background changes and the wavelength. 
There is, however, no further restriction on the velocity amplitude and therefore it is allowed 
to be of the same order as the background flow in contrast to weakly nonlinear theory where the amplitude itself is the expansion parameter.
Hence, we may call Grimshaw's modulation equations a strongly nonlinear wave theory.

The notion of ``strongly nonlinear'' is stipulated in the following sense. 
A superposition of solutions from the class of strongly nonlinear waves 
does not lead to small errors when inserted into the governing equations (the Euler or Navier-Stokes equations)
as it is the case for the weakly nonlinear theory.
Instead such a superposition generates terms by wave-wave interaction of order unity in the asymptotic limit.
However, this notion does not necessarily imply that the nonlinearities in the governing equation, i.e. the advection terms,
are bigger than the pressure gradient term, say, in terms of non-dimensional analysis. 
In fact, the asymptotic expansion of Grimshaw's theory is such that the perturbation advection terms vanish due to the solenoidality of the wind field.
%In this sense the notion of strong nonlinearity simply has to be seen in contrast to weakly nonlinear wave theory.
Some theoretical investigations on strongly nonlinear effects and their implications with respect to observations and modeling 
were performed in \cite{Schlutow2018,Schlutow2019}.

The objective of this paper is to investigate the interaction of strongly nonlinear waves with a very thin shear layer.
This renders a common situation in the real atmosphere, e.g., when a mountain wave meets the tropospheric jet and gets refracted \citep{Ehard2017}.
In the setting of linear theory, this situation was studied in the seminal work \cite{Eliassen1961} using a layered model.
They approximated the height-dependent background by piecewise constant functions. 
This idea was advanced in \cite{Putz2019} for linear wave packets.
We will adopt the idea of a layered background atmosphere but for strongly nonlinear waves. 

Our main results are necessary and sufficient criteria for the stability of a refracted non-hydrostatic wave in a two-layered background. 
If the mean-flow horizontal wind in the lower layer is stronger than the wind in the upper layer, 
than the wave is stable with respect to the modulation equations.
If, however, the upper-layer mean-flow horizontal wind is stronger, which typically occurs for a gravity wave entering the jet, 
the refracted wave becomes unstable if both mean-flow horizontal winds meet particular upper bounds 
and if the amplitude of the wave is sufficiently large. 

This work is structured as follows. In section \ref{sec:model}, we will introduce Grimshaw's modulation equations as our governing equations. 
Section \ref{sec:solution} will be dedicated to the strongly nonlinear wave refracted at a discontinuity as a particular solution to the modulation equations.
The stability of the refracted wave will be analyzed in section \ref{sec:stability}. We will summarize the results in section \ref{sec:summary} 
and offer some concluding remarks in section \ref{sec:conlcusion}.

In the second part, we will investigate the wave solution and the stability results numerically by means of Large Eddy Simulations.

%The interaction of weakly nonlinear waves with a jet were analyzed numerically by \citep{Boloni2016}.

\section{The model equations}
\label{sec:model}

We consider vertically modulated, horizontally periodic, non-hydrostatic, 
and strongly nonlinear gravity waves in the unbounded $x$-$z$-plane 
which we model by Grimshaw's modulation equations \citep{Grimshaw1974,Achatz2010,Schlutow2017b},
\begin{subequations}
\begin{align}
	\label{eq:wn_evo}
	&\frac{\partial k_z}{\partial t}+\frac{\partial \omega}{\partial z}=0\\
	\label{eq:wad_evo}
	&\rho\frac{\partial a}{\partial t}+\frac{\partial}{\partial z}(\hat\omega'\rho a)=0\\
	\label{eq:mf_evo}
	&\rho\frac{\partial u}{\partial t}+\frac{\partial}{\partial z}(\hat\omega'k_x\rho a)=0.
\end{align}
\end{subequations}
This coupled set of equations governs the evolution of vertical wavenumber $k_z$, 
wave action density $\rho a$, and mean-flow horizontal wind $u$. 
The horizontal wavenumber $k_x=K_x$ is a constant being, without loss of generality, positive. 
The Extrinsic frequency $\omega=\hat\omega+k_xu$ is determined by the Doppler-shifted intrinsic frequency 
which depends on wavenumber due to the dispersion relation for non-hydrostatic waves,
\begin{align}
	\label{eq:dispersion}
	\hat{\omega}=\frac{Nk_x}{|\pmb{k}|},\quad|\pmb{k}|=\sqrt{k_x^2+k_z^2}.
\end{align}
Its derivative with respect to the vertical wavenumber,
\begin{align}
	\hat{\omega}'=-\frac{Nk_xk_z}{|\pmb{k}|^3},
\end{align}
represents the vertical linear group velocity.
Two coefficients appear representing the background state of the atmosphere, the Brunt-V\"ais\"al\"a frequency $N$ 
and the density $\rho$ which are generally functions of altitude, $z$.
However, we apply the Boussinesq assumption such that
\begin{align}
	\rho,N=\mathrm{const}.
\end{align}

In Whitham’s modulation theory \citep{Whitham1974a}, equation \eqref{eq:wn_evo} represents conservation of waves.
The second equation \eqref{eq:wad_evo} yields conservation of wave action. 
Finally, \eqref{eq:mf_evo} describes the acceleration of mean-flow horizontal momentum 
due to the convergence of the flux of horizontal pseudo-momentum $k_x\rho a$.

%\emph{
%In this study we are interested in the interaction of a strongly nonlinear wave with a very thin shear layer. 
%The modulation equations are formulated on a spatial scale comparable to the scale height, approximately 10\,km, 
%for waves with much smaller local wavelengths. 
%If we assume that the thickness of the shear layer is of similar order of magnitude as the wavelength, 
%then the change in mean-flow horizontal velocity appears as a discontinuous jump in the modulation equations. 
%In particular, the scale of change is much smaller than the scale on which the background density varies 
%and therefore it is appropriate to assume}

%which corresponds to an isothermal Boussinesq atmosphere.

\section{The refracted wave solution}
\label{sec:solution}

In this section, we construct an analytical solution to the modulation equations reminiscent of a typical mountain wave 
that encounters a jet at a certain height---e.g. the tropospheric or mesospheric jet---and experiences refraction.
This is a classical situation from linear wave theory.
In contrast to linear theory however, we will see that the wave drags the background flow by direct nonlinear interaction
resulting in a self-induced mean flow and Doppler shifting.

When multiplying $k_x$ to \eqref{eq:wad_evo}, then subtracting \eqref{eq:mf_evo} and integrating we find
\begin{align}
	\label{eq:momentum}
	u(z,t)=k_xa(z,t)+\bar{U}(z).
\end{align}
We also find $\bar{U}$ as an integration constant that 
can be identified as the horizontal background wind, the wind without the occurrence of a wave ($a=0$),
and $u$ is really the mean-flow horizontal wind.

\begin{figure}
	\begin{center}
		\input{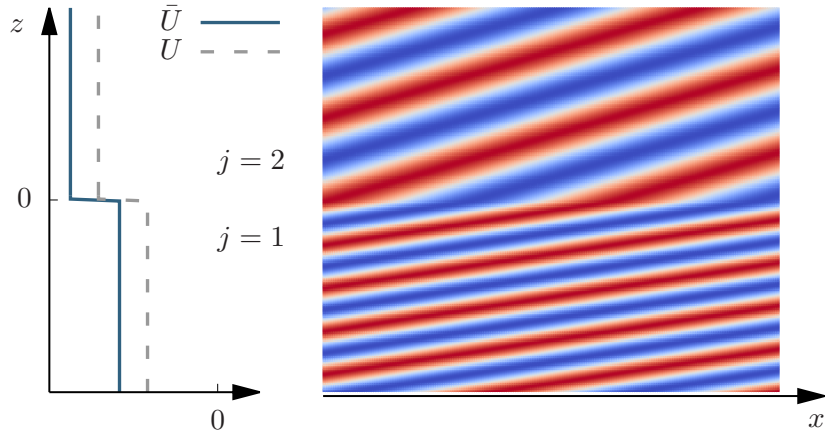}	
	\end{center}
	\caption{Sketch of the refracted wave solution}
	\label{fig:solution}
\end{figure}	

Let us consider a vertically sheared background flow being piecewise constant 
with the discontinuity at $z=0$,
\begin{align}
	\bar{U}(z)=
	\begin{cases}
		\bar{U}_1,\quad z<0,\\
		\bar{U}_2,\quad z>0. 
	\end{cases}
\end{align}
Then a stationary solution is given by
\begin{align}
	(k_z(z,t),\,a(z,t))^\mathrm{T}=(K_z(z),\,A(z))^\mathrm{T}=\begin{cases}
		(K_{z,1},\,A_{1})^\mathrm{T},\quad z<0,\\
		(K_{z,2},\,A_{2})^\mathrm{T},\quad z>0.
	\end{cases}
\end{align}
The mean-flow horizontal wind really becomes a diagnostic variable due to \eqref{eq:momentum},
\begin{align}
	u(z,t)=U(z)=\begin{cases}
		U_1,\quad z<0,\\
		U_2,\quad z>0, 
	\end{cases}\quad
	U_j=K_xA_j+\bar{U}_j,\quad j=1,2.
\end{align}
Here and further on in this investigation, $j=1$ denotes the layer below and $j=2$ the layer above the discontinuity at $z=0$.
The constant pieces are determined by integration of \eqref{eq:wn_evo} and \eqref{eq:wad_evo},
\begin{subequations}
\begin{align}
	\label{eq:omega}
	\hat{\omega}(K_{z,j})+K_xU_j&=\mathrm{constant},\\
	\label{eq:waveactionflux}
	\hat{\omega}'(K_{z,j})A_j&=\mathrm{constant}.
\end{align}
\end{subequations}
The integration constant in \eqref{eq:omega} must be zero as the extrinsic frequency is readily the derivative of the phase function with respect to time.
Therefore, a stationary phase requires vanishing frequency.
By means of this argument and \eqref{eq:dispersion}, \eqref{eq:omega} can be rearranged to obtain
\begin{align}
	\label{eq:k_z}
	K_{z,j}=-\sqrt{\frac{N^2}{U_j^2}-K_x^2}.
\end{align}
Equation \eqref{eq:waveactionflux} provides an interesting physical implication: the vertical wave action flux is invariant throughout the layered atmosphere. 
And in particular, it is constant crossing the interface of the two layers.
In order to ensure internal waves we must find
\begin{align}
	\label{eq:evanescent}
	J_j\equiv\frac{N^2}{K_x^2U_j^2}>1\quad\text{for all}\quad j=1,2.
\end{align}
Otherwise the discriminant in \eqref{eq:k_z} is negative corresponding to evanescent or rather external waves.
We have introduced a new parameter $J_j$ which will become important in the following analysis.
Please note that the mean-flow horizontal wind $u$ has become a diagnostic variable due to \eqref{eq:momentum}.
Let us concatenate the prognostic variables of the wave solution into a vector,
\begin{align}
	\pmb{P}(z)=(K_z,A)^\mathrm{T}.
\end{align}
A sketch of the wave which would be generated by this solution is plotted in figure \ref{fig:solution}.
Due to the larger background velocity in the upper layer, the phase lines are more inclined than in the lower layer.
%
%\begin{figure}
%\centering
%\includegraphics[scale=0.5,angle=270]{sketch_of_refracted_wave.pdf} 
%\caption{Sketch of the refracted wave solution}
%\end{figure}
%

\section{Modulational stability of the refracted wave}
\label{sec:stability}

Linearizing the governing equations reduced by \eqref{eq:momentum} around the stationary solution $\pmb{P}$ combined with an ansatz for the perturbation
\begin{align}
	\pmb{p}(z,t)=\tilde{\pmb{p}}(z)e^{\lambda t},\quad\pmb{p}(z,t)=(k_z,a)^\mathrm{T}
\end{align}
yields an eigenvalue problem $(\mathcal{L}-\lambda)\tilde{\pmb{p}}=0$
for a linear differential operator $\mathcal{L}$
which can be reformulated as an ordinary differential equation
\begin{align}
	\label{eq:lin_ode}
	\frac{d\tilde{\pmb{p}}(z)}{dz}=\mathbf{B}(z,\lambda)\,\tilde{\pmb{p}}(z).
\end{align}	
The coefficient matrix contains the refracted wave solution and is therefore piecewise constant in $z$,
\begin{align}
	\mathbf{B}(z,\lambda)=
	\begin{cases}
		\mathbf{B}_1,\quad z<0,\\
		\mathbf{B}_2,\quad z>0.
	\end{cases}
\end{align}
where
\begin{align}
	\label{eq:ode_matrix}
	\mathbf{B}_j&=\frac{\lambda}{C_j^2-K_x^2A_jH_j}
	\begin{pmatrix}
		-C_j& K_x^2\\
		A_jH_j& -C_j
	\end{pmatrix},\\
	C_j&=\hat{\omega}'(K_{z,j}),\\
	\label{eq:h}
	H_j&=\hat{\omega}''(K_{z,j})=NK_x\frac{2K_{z,j}^2-K_x^2}{|\pmb{K}_j|^5}.
\end{align}
To decide about the stability of our refracted wave solution,
we will investigate the spectrum of the linear operator posed by the linearized governing equations.
Roughly speaking, the spectrum is the set of all $\lambda\in\mathbb{C}$ for which our operator is not invertible.
If the spectrum is contained on the left hand side of the complex plane, 
i.e. all $\Re(\lambda)\leq 0$, then we can say the solution is spectrally stable and unstable otherwise.
The spectrum consists of two qualitatively different subsets defined by its Fredholm properties:
the essential (continuous) spectrum and the point (matrix-like) spectrum. 
We elaborated on this definition in terms of Fredholm operators 
in the context of strongly nonlinear gravity waves in \cite{Schlutow2018}.
For the sake of brevity, the details are not repeated in this paper.
We refer to the book of \cite{Kapitula2013} for a detailed introduction and precise definitions.

\subsection{The essential spectrum}

The essential spectrum is given by the hyperbolicity of the matrix \eqref{eq:ode_matrix} \cite{Sandstede2002,Kapitula2013}.
A matrix is said to be hyperbolic if all its eigenvalues have non-zero real part.
The boundaries of the open regions in the complex plane which comprise the essential spectrum coincide exactly with 
\eqref{eq:ode_matrix} being not hyperbolic.
Therefore, the ansatz
\begin{align}
	\det(\mathbf{B}_j(\lambda)-i\mu)=0,\quad\mu\in\mathbb{R}
\end{align}
leads to 
\begin{align}
	\lambda_j^\pm(\mu)=-iC_j\,\mu\pm iK_x\sqrt{A_jH_j}\,\mu
\end{align}
which represents four parameterized curves in the complex plane enclosing the essential spectrum.
In particular, we will find unstable essential spectrum if one of the boundary curves enters into the right half of the complex plane
as either left or right from the curve we will find essential spectrum.
This occurs when $H_j<0$ which corresponds to the classical modulational instability criterion \citep{Grimshaw1977,Schlutow2018,Schlutow2019}. 
The instability growth rate is then $K_x\sqrt{A_j|H_j|}|\mu|$ which happens to be unbounded in $\mu$ or in other words
arbitrarily large $\mu$ results in arbitrarily large instability growth which would render the problem unphysical and mathematically ill-posed.
The classical modulational instability criterion can be rewritten by insertion of \eqref{eq:k_z} into \eqref{eq:h} as
\begin{align}
	J_j>\frac{3}{2}
\end{align}
providing us with a new lower bound greater than the criterion for non-evanescence in \eqref{eq:evanescent}.

\subsection{The point spectrum}

In the following, we assume a stable essential spectrum, so $H_1>0$ and $H_2>0$.
The point spectrum consists of eigenvalues to which the corresponding eigenfunctions belong.
Eigenfunctions must be of the form
\begin{align}
	\label{eq:ansatz_ef}
	\tilde{\pmb{p}}=
	\begin{cases}
		\tilde{\pmb{p}}_1,\quad z<0\\
		\tilde{\pmb{p}}_2,\quad z>0\\
	\end{cases},\quad
	\tilde{\pmb{p}}_j=\hat{\pmb{p}}_je^{\sigma_j z}+\mathrm{cc}
\end{align}
with cc denoting the complex conjugate.
We obtain the spatial eigenvalues $\sigma_j$ by the solvability condition of the emerging algebraic system
when inserting the ansatz \eqref{eq:ansatz_ef} into \eqref{eq:lin_ode},
\begin{align}
	\label{eq:spatial_ev}
	\sigma_j^\pm=\frac{\lambda}{\pm K_x\sqrt{A_jH_j}-C_j}.
\end{align}
Please note that $\pm$ indicates two linearly independent solutions.
Special care has to be taken for $\pm K_x\sqrt{A_jH_j}=C_j$.
In this very particular case, $\lambda=0$ is the only eigenvalue.
Associated basis vectors to the eigenvalues \eqref{eq:spatial_ev} may be written as
\begin{align}
	\pmb{q}_j^\pm=(-K_x^2\sigma_j^\pm,\,\lambda+C_j\sigma_j^\pm)^\mathrm{T}.
\end{align}
In terms of the basis vectors, the eigenvectors are given by superposition, i.e. linear combination, 
\begin{align}
	\tilde{\pmb{p}}_j=b_j^+\pmb{q}_j^+e^{\sigma_j^+z}+b_j^-\pmb{q}_j^-e^{\sigma_j^-z}+\mathrm{cc}.
\end{align}
To be an eigenfunction at least one of the coefficients $b_j^\pm$ must differ from zero. 
Because if all coefficients vanish, then the solution is trivial and not an eigenfunction.
The eigenfunctions must obey three constraints. 
\begin{enumerate}
	\item \label{con:1} In the limit $z\rightarrow-\infty$, the eigenfunctions must vanish.
	\item \label{con:2} In the limit $z\rightarrow+\infty$, the eigenfunctions must vanish.
	\item \label{con:3} At the discontinuity $z=0$, the eigenfunctions need to be continuous.
\end{enumerate}
These three constraints constitute perturbations of finite energy 
as well as continuous pressure and vertical wind velocity at the discontinuity of the backround horizontal wind.
%
%\begin{figure}
%\centering
%\includegraphics[scale=0.5,angle=270]{Sketch_perturbation.pdf} 
%\caption{Sketch of the form of the perturbation}
%\end{figure}
%
\begin{figure}
	\begin{center}
		\input{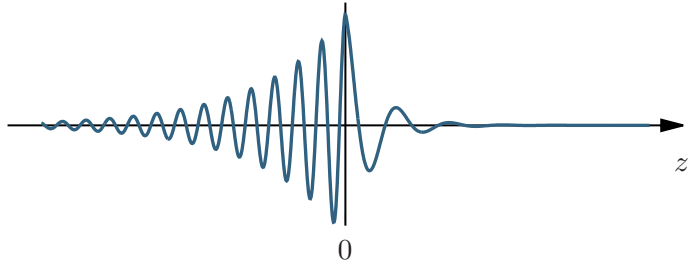}	
	\end{center}
	\caption{Sketch of the perturbation}
	\label{fig:perturbation}
\end{figure}	
Figure \ref{fig:perturbation} presents a sketch of the form of the eigenfunctions in line with the constraints \ref{con:1}--\ref{con:3}.
They oscillate and decay away from the jump.

Let us fix in the following $\Re(\lambda)>0$ and by this we ask for unstable eigenvalues. 
We consider \eqref{eq:spatial_ev} and distinguish the four different cases that may occur regarding its sign.
\begin{itemize}
	\item If $K_x\sqrt{A_1H_1}>C_1$, then $\Re(\sigma_1^+)>0$ and $\Re(\sigma_1^-)<0$. 
	To meet constraint \ref{con:1}, $b_1^-=0$ must therefore be true
	whereas $b_1^+\neq 0$ is admissible.
	\item If  $K_x\sqrt{A_1H_1}<C_1$, then $\Re(\sigma_1^+)<0$ and $\Re(\sigma_1^-)<0$.
	To meet constraint \ref{con:1}, ${b_1^+=b_1^-=0}$ must therefore be true.
	\item If $K_x\sqrt{A_2H_2}>C_2$, then $\Re(\sigma_2^+)>0$ and  $\Re(\sigma_2^-)<0$.
	To meet constraint \ref{con:2}, $b_2^+= 0$ must therefore be true
	whereas $b_2^-\neq 0$ is admissible.
	\item If $K_x\sqrt{A_2H_2}<C_2$, then $\Re(\sigma_2^+)<0$ and $\Re(\sigma_2^-)<0$.
	Hence, constraint \ref{con:2} allows for $b_2^+\neq 0$ and $b_2^-\neq 0$.
\end{itemize}
Next, we investigate the continuity constraint 3, i.e. $\tilde{\pmb{p}}_1=\tilde{\pmb{p}}_2$ at $z=0$ which yields
\begin{align}
	\label{eq:continuity}
	b_1^+\pmb{q}_1^+=b_2^+\pmb{q}_2^++b_2^-\pmb{q}_2^-.
\end{align}
Note that we already anticipated that in all cases $b_1^-=0$.
Given $b_1^+$ we can solve \eqref{eq:continuity} for
\begin{subequations}
\label{eq:b_2}
\begin{align}
	\label{eq:b_2a}
	b_2^+=\frac{1}{2}\frac{+K_x\sqrt{A_2H_2}-C_2}{K_x\sqrt{A_1H_1}-C_1}\left(1+\sqrt{\frac{H_1}{H_2}}\right)b_1^+,\\
	\label{eq:b_2b}
	b_2^-=\frac{1}{2}\frac{-K_x\sqrt{A_2H_2}-C_2}{K_x\sqrt{A_1H_1}-C_1}\left(1-\sqrt{\frac{H_1}{H_2}}\right)b_1^+.
\end{align}
\end{subequations}
We read immediately from \eqref{eq:b_2} that $b_2^+=b_2^-=0$ if $b_1^+=0$ 
which occurs when $K_x\sqrt{A_1H_1}<C_1$ according to constraint \ref{con:1}.
The latter renders therefore a sufficient condition for stability as, if it is true, 
there exist only trivial solution for any $\Re(\lambda)>0$ 
and hence it cannot be an eigenvalue.
Furthermore, $b_2^+=0$ fulfilling constraint \ref{con:2} is only possible if $b_1^+=0$ due to \eqref{eq:b_2a}, but then \eqref{eq:b_2b} yields also $b_2^-=0$.
And again we are left with the trivial solution.
By the same argument, $K_x\sqrt{A_2H_2}>C_2$ is hence a sufficient condition for stability.
In conclusion, there are nontrivial solutions, which happen to be unstable, if 
\begin{subequations}
\label{eq:ess_crit}
\begin{align}
	K_x\sqrt{A_2H_2}&<C_2,\\
	K_x\sqrt{A_1H_1}&>C_1.
\end{align}
\end{subequations}

\section{Summary}
\label{sec:summary}

In this section, we summarize and reformulate the results of the previous sections.
We may rewrite our criterion from the point spectrum in terms 
of the saturation amplitudes of the waves in each layer $\alpha_j$ 
which relates to the wave action density by
\begin{align}
	\label{eq:ampl_to_wad}
	A_j=\frac{N^2}{2\hat{\omega}^2(K_{z,j})}\alpha_j^2.
\end{align}
Waves become unstable due to static instabilities if $\alpha_j\ge 1$ \cite{Boloni2016}.
Thus, by inserting \eqref{eq:ampl_to_wad} and \eqref{eq:evanescent} into the criteria \eqref{eq:ess_crit} 
and rearranging, the refracted wave is stationary stable but unstable 
due to perturbations from the point spectrum if
\begin{subequations}
\label{eq:b_alpha}
\begin{align}
	\label{eq:lb_alpha_1}
	1&>\alpha_1^2>\frac{2}{J_1}\frac{(J_1-1)^2}{2J_1-3},\\
	\label{eq:ub_alpha_2}
	\frac{2}{J_2}\frac{(J_2-1)^2}{2J_2-3}&>\alpha_2^2>0.
\end{align}
\end{subequations}
To have unity as the upper bound for the right hand side in \eqref{eq:lb_alpha_1} 
implies that $J_1\in(2,\infty)$.
It represents a particularly narrowing bound on the amplitude in the lower layer as,
\begin{align}
	\min_{J_1\in(2,\infty)}\frac{2}{J_1}\frac{(J_1-1)^2}{2J_1-3}=\frac{8}{9}.
\end{align}
The minimum is assumed at $J_1=3$, so at best $1>\alpha_1>\sqrt{8/9}\approx 0.94$.
Zero as the lower bound for the left hand side in \eqref{eq:ub_alpha_2}
is met if $J>3/2$ which is the same bound from the stable essential spectrum.

Utilizing the fact that the vertical wave action flux is invariant across the interface (cf. \eqref{eq:waveactionflux}), 
we find $C_1A_1=C_2A_2$. We combine our finding with \eqref{eq:ampl_to_wad} and obtain
\begin{align}
	\alpha_1^2=\sqrt{\frac{J_1-1}{J_2-1}}\frac{J_1}{J_2}\alpha_2^2
\end{align}
which we substitute in \eqref{eq:b_alpha}. The resulting inequality reads
\begin{align}
	\frac{1}{J_2^2}\frac{(J_2-1)^{3/2}}{2J_2-3}>\frac{1}{J_1^2}\frac{(J_1-1)^{3/2}}{2J_1-3}
\end{align}
being the same as $g(J_2)>g(J_1)$. 
Notice that the function $g$ is continuous as well as monotonically decreasing in $(3/2,\infty)$
and hence
\begin{align}
	J_2<J_1
\end{align}
which implies due to \eqref{eq:evanescent} that $|U_1|<|U_2|$.

In conclusion, a refracted gravity wave, that is statically stable, non-evanescent, non-hydrostatic 
and features a stable essential spectrum of the linearized modulation equations,
becomes unstable due to perturbations from the point spectrum if
\begin{subequations}
\begin{align}
	|U_1|<\frac{1}{\sqrt{2}}\frac{N}{K_x},\quad|U_2|<\sqrt{\frac{2}{3}}\frac{N}{K_x},\quad|U_1|<|U_2|
\end{align}
and
\begin{align}
	\alpha_1^2>\frac{2}{J_1}\frac{(J_1-1)^2}{2J_1-3}.
\end{align}
\end{subequations}

\section{Discussion}
\label{sec:conlcusion}

On the one hand, the constraint on the saturation amplitude in the lower layer for an emerging instability is extreme. It is at best 94\,\% or higher.
On the other hand, the conditions for the mean-flow horizontal wind favorable for instability are fairly common. 
A typical situation where this type of instability may occur is the lower edge of the atmospheric jets where a mountain wave enters a strong shear zone from below and gets refracted into the jet. It is not unlikely that such a wave has large amplitude due to anelastic amplification.

The perturbation affects not only the wave properties but also the mean-flow horizontal wind as a diagnostic variable due to \eqref{eq:momentum}.
Also we observe that perturbations decay exponentially away from the discontinuity of the background wind (cf. figure \ref{fig:perturbation} for illustration).
We conclude therefore that a point-spectrum instability is localized at the jump. 
Due to these observations we can suspect an attenuation of the sharp gradient in the background horizontal wind by an instability from the point spectrum.
In conclusion, strongly nonlinear gravity waves may serve as a mitigation of strongly sheared flows.

Moreover, Grimshaw's modulation equations, which provide the basis for our analysis, were derived from the Euler equations using WKB theory.  
A crucial assumption in the derivation is that the mean-flow varies on a larger scale than a typical wavelength. 
By assuming vertically piecewise constant mean-flow horizontal wind, we potentially infringe on this WKB constraint.
The range of validity of our results will be tested by means of numerical simulations of the Euler equations in a successive paper.

	\section*{Acknowledgments}
	This research was supported by the German Research Foundation (DFG) through the Research Unit FOR 1898 (MS-GWaves),
through grants KL 611/25-2, AC 71/10-2 and BO 5071/1-2.	
	
%	\appendix	

%	\input{Appendix.tex}

% 	\clearpage

%	\printbibliography[heading=bibintoc]
	\bibliographystyle{abbrvnat}
	\bibliography{library.bib}

\begin{thebibliography}{27}
\providecommand{\natexlab}[1]{#1}
\providecommand{\url}[1]{\texttt{#1}}
\expandafter\ifx\csname urlstyle\endcsname\relax
  \providecommand{\doi}[1]{doi: #1}\else
  \providecommand{\doi}{doi: \begingroup \urlstyle{rm}\Url}\fi

\bibitem[Achatz et~al.(2010)Achatz, Klein, and Senf]{Achatz2010}
U.~Achatz, R.~Klein, and F.~Senf.
\newblock {Gravity waves, scale asymptotics and the pseudo-incompressible
  equations}.
\newblock \emph{Journal of Fluid Mechanics}, 663:\penalty0 120--147, 2010.
\newblock \doi{10.1017/S0022112010003411}.

\bibitem[Becker(2004)]{Becker2004}
E.~Becker.
\newblock {Direct heating rates associated with gravity wave saturation}.
\newblock \emph{Journal of Atmospheric and Solar-Terrestrial Physics},
  66\penalty0 (6-9):\penalty0 683--696, 2004.
\newblock \doi{10.1016/j.jastp.2004.01.019}.

\bibitem[Becker(2012)]{Becker2012}
E.~Becker.
\newblock {Dynamical control of the middle atmosphere}.
\newblock \emph{Space Science Reviews}, 168\penalty0 (1-4):\penalty0 283--314,
  2012.
\newblock \doi{10.1007/s11214-011-9841-5}.

\bibitem[B{\"{o}}l{\"{o}}ni et~al.(2016)B{\"{o}}l{\"{o}}ni, Ribstein,
  Muraschko, Sgoff, Wei, and Achatz]{Boloni2016}
G.~B{\"{o}}l{\"{o}}ni, B.~Ribstein, J.~Muraschko, C.~Sgoff, J.~Wei, and
  U.~Achatz.
\newblock {The interaction between atmospheric gravity waves and large-scale
  flows: An efficient description beyond the nonacceleration paradigm}.
\newblock \emph{Journal of the Atmospheric Sciences}, 73\penalty0
  (12):\penalty0 4833--4852, aug 2016.
\newblock \doi{10.1175/JAS-D-16-0069.1}.

\bibitem[Bretherton(1966)]{Bretherton1966}
F.~Bretherton.
\newblock {The propagation of groups of internal gravity waves in a shear
  flow}.
\newblock \emph{Quarterly Journal of the Royal Meteorological Society},
  92\penalty0 (394):\penalty0 466--480, 1966.
\newblock \doi{10.1002/qj.49709239403}.

\bibitem[Dunkerton(1978)]{Dunkerton1978}
T.~Dunkerton.
\newblock {On the Mean Meridional Mass Motions of the Stratosphere and
  Mesosphere}.
\newblock \emph{Journal of the Atmospheric Sciences}, 35\penalty0
  (12):\penalty0 2325--2333, dec 1978.
\newblock \doi{10.1175/1520-0469(1978)035<2325:OTMMMM>2.0.CO;2}.

\bibitem[Ehard et~al.(2017)Ehard, Kaifler, D{\"{o}}rnbrack, Preusse, Eckermann,
  Bramberger, Gisinger, Kaifler, Liley, Wagner, and Rapp]{Ehard2017}
B.~Ehard, B.~Kaifler, A.~D{\"{o}}rnbrack, P.~Preusse, S.~D. Eckermann,
  M.~Bramberger, S.~Gisinger, N.~Kaifler, B.~Liley, J.~Wagner, and M.~Rapp.
\newblock {Horizontal propagation of large-amplitude mountain waves into the
  polar night jet}.
\newblock \emph{Journal of Geophysical Research}, 122\penalty0 (3):\penalty0
  1423--1436, 2017.
\newblock \doi{10.1002/2016JD025621}.

\bibitem[Eliassen and Palm(1961)]{Eliassen1961}
A.~Eliassen and E.~Palm.
\newblock {On the Transfer of Energy in Stationary Mountain Waves}.
\newblock \emph{Geofysiske Publikasjoner}, 22:\penalty0 1--23, 1961.

\bibitem[Fritts and Alexander(2003)]{Fritts2003}
D.~C. Fritts and M.~J. Alexander.
\newblock {Gravity wave dynamics and effects in the middle atmosphere}.
\newblock \emph{Reviews of Geophysics}, 41\penalty0 (1):\penalty0 1003, 2003.
\newblock \doi{10.1029/2001RG000106}.

\bibitem[Fritts et~al.(2016)Fritts, Smith, Taylor, Doyle, Eckermann,
  D{\"{o}}rnbrack, Rapp, Williams, Pautet, Bossert, Criddle, Reynolds,
  Reinecke, Uddstrom, Revell, Turner, Kaifler, Wagner, Mixa, Kruse, Nugent,
  Watson, Gisinger, Smith, Lieberman, Laughman, Moore, Brown, Haggerty,
  Rockwell, Stossmeister, Williams, Hernandez, Murphy, Klekociuk, Reid, and
  Ma]{Fritts2016}
D.~C. Fritts, R.~B. Smith, M.~J. Taylor, J.~D. Doyle, S.~D. Eckermann,
  A.~D{\"{o}}rnbrack, M.~Rapp, B.~P. Williams, P.-D. Pautet, K.~Bossert, N.~R.
  Criddle, C.~A. Reynolds, P.~A. Reinecke, M.~Uddstrom, M.~J. Revell,
  R.~Turner, B.~Kaifler, J.~S. Wagner, T.~Mixa, C.~G. Kruse, A.~D. Nugent,
  C.~D. Watson, S.~Gisinger, S.~M. Smith, R.~S. Lieberman, B.~Laughman, J.~J.
  Moore, W.~O. Brown, J.~A. Haggerty, A.~Rockwell, G.~J. Stossmeister, S.~F.
  Williams, G.~Hernandez, D.~J. Murphy, A.~R. Klekociuk, I.~M. Reid, and J.~Ma.
\newblock {The Deep Propagating Gravity Wave Experiment (DEEPWAVE): An Airborne
  and Ground-Based Exploration of Gravity Wave Propagation and Effects from
  Their Sources throughout the Lower and Middle Atmosphere}.
\newblock \emph{Bulletin of the American Meteorological Society}, 97\penalty0
  (3):\penalty0 425--453, mar 2016.
\newblock \doi{10.1175/BAMS-D-14-00269.1}.

\bibitem[Fritts et~al.(2018)Fritts, Vosper, Williams, Bossert, Plane, Taylor,
  Pautet, Eckermann, Kruse, Smith, D{\"{o}}rnbrack, Rapp, Mixa, Reid, and
  Murphy]{Fritts2018}
D.~C. Fritts, S.~B. Vosper, B.~P. Williams, K.~Bossert, J.~M.~C. Plane, M.~J.
  Taylor, P.-D. Pautet, S.~D. Eckermann, C.~G. Kruse, R.~B. Smith,
  A.~D{\"{o}}rnbrack, M.~Rapp, T.~Mixa, I.~M. Reid, and D.~J. Murphy.
\newblock {Large-Amplitude Mountain Waves in the Mesosphere Accompanying Weak
  Cross-Mountain Flow During DEEPWAVE Research Flight RF22}.
\newblock \emph{Journal of Geophysical Research: Atmospheres}, 123\penalty0
  (18):\penalty0 9992--10022, sep 2018.
\newblock \doi{10.1029/2017JD028250}.

\bibitem[Fritts et~al.(2019)Fritts, Wang, Taylor, Pautet, Criddle, Kaifler,
  Eckermann, and Liley]{Fritts2019}
D.~C. Fritts, L.~Wang, M.~J. Taylor, P.~D. Pautet, N.~R. Criddle, B.~Kaifler,
  S.~D. Eckermann, and B.~Liley.
\newblock {Large-Amplitude Mountain Waves in the Mesosphere Observed on 21 June
  2014 During DEEPWAVE: 2. Nonlinear Dynamics, Wave Breaking, and
  Instabilities}.
\newblock \emph{Journal of Geophysical Research: Atmospheres}, 124\penalty0
  (17-18):\penalty0 10006--10032, sep 2019.
\newblock \doi{10.1029/2019JD030899}.

\bibitem[Grimshaw(1972)]{Grimshaw1972}
R.~Grimshaw.
\newblock {Nonlinear internal gravity waves in a slowly varying medium}.
\newblock \emph{Journal of Fluid Mechanics}, 54\penalty0 (2):\penalty0
  193--207, 1972.
\newblock \doi{10.1017/S0022112072000631}.

\bibitem[Grimshaw(1974)]{Grimshaw1974}
R.~Grimshaw.
\newblock {Internal gravity waves in a slowly varying, dissipative medium}.
\newblock \emph{Geophysical Fluid Dynamics}, 6:\penalty0 131--148, 1974.
\newblock \doi{10.1080/03091927409365792}.

\bibitem[Grimshaw(1977)]{Grimshaw1977}
R.~Grimshaw.
\newblock {The Modulation of an Internal Gravity-Wave Packet, and the Resonance
  with the Mean Motion}.
\newblock \emph{Studies in Applied Mathematics}, 56\penalty0 (3):\penalty0
  241--266, jun 1977.
\newblock \doi{10.1002/sapm1977563241}.

\bibitem[Kapitula and Promislow(2013)]{Kapitula2013}
T.~Kapitula and K.~Promislow.
\newblock \emph{{Spectral and Dynamical Stability of Nonlinear Waves}}, volume
  185 of \emph{Applied Mathematical Sciences}.
\newblock Springer New York, New York, NY, 2013.
\newblock ISBN 978-1-4614-6994-0.
\newblock \doi{10.1007/978-1-4614-6995-7}.

\bibitem[Lindzen(1981)]{Lindzen1981}
R.~S. Lindzen.
\newblock {Turbulence and stress owing to gravity wave and tidal breakdown}.
\newblock \emph{Journal of Geophysical Research}, 86\penalty0 (C10):\penalty0
  9707--9714, oct 1981.
\newblock \doi{10.1029/JC086iC10p09707}.

\bibitem[P{\"{u}}tz et~al.(2019)P{\"{u}}tz, Schlutow, and Klein]{Putz2019}
C.~P{\"{u}}tz, M.~Schlutow, and R.~Klein.
\newblock {Initiation of ray tracing models: evolution of small-amplitude
  gravity wave packets in non-uniform background}.
\newblock \emph{Theoretical and Computational Fluid Dynamics}, 33\penalty0
  (5):\penalty0 509--535, oct 2019.
\newblock \doi{10.1007/s00162-019-00504-z}.

\bibitem[Sandstede(2002)]{Sandstede2002}
B.~Sandstede.
\newblock {Stability of travelling waves}.
\newblock In B.~Fiedler, editor, \emph{Handbook of dynamical systems},
  volume~2, pages 983--1055. Gulf Professional Publishing, 2002.
\newblock ISBN 978-0444826695.

\bibitem[Schlutow(2019)]{Schlutow2019}
M.~Schlutow.
\newblock {Modulational Stability of Nonlinear Saturated Gravity Waves}.
\newblock \emph{Journal of the Atmospheric Sciences}, 76\penalty0
  (11):\penalty0 3327--3336, nov 2019.
\newblock \doi{10.1175/JAS-D-19-0065.1}.

\bibitem[Schlutow et~al.(2014)Schlutow, Becker, and
  K{\"{o}}rnich]{Schlutow2014}
M.~Schlutow, E.~Becker, and H.~K{\"{o}}rnich.
\newblock {Positive definite and mass conserving tracer transport in spectral
  GCMs}.
\newblock \emph{Journal of Geophysical Research: Atmospheres}, 119\penalty0
  (20):\penalty0 11,511--562,577, 2014.
\newblock \doi{10.1002/2014JD021661}.

\bibitem[Schlutow et~al.(2017)Schlutow, Klein, and Achatz]{Schlutow2017b}
M.~Schlutow, R.~Klein, and U.~Achatz.
\newblock {Finite-amplitude gravity waves in the atmosphere: travelling wave
  solutions}.
\newblock \emph{Journal of Fluid Mechanics}, 826:\penalty0 1034--1065, sep
  2017.
\newblock \doi{10.1017/jfm.2017.459}.

\bibitem[Schlutow et~al.(2019)Schlutow, Wahl{\'{e}}n, and Birken]{Schlutow2018}
M.~Schlutow, E.~Wahl{\'{e}}n, and P.~Birken.
\newblock {Spectral stability of nonlinear gravity waves in the atmosphere}.
\newblock \emph{Mathematics of Climate and Weather Forecasting}, 5\penalty0
  (1):\penalty0 12--33, 2019.
\newblock \doi{10.1515/mcwf-2019-0002}.

\bibitem[Sutherland(2001)]{Sutherland2001}
B.~R. Sutherland.
\newblock {Finite-amplitude internal wavepacket dispersion and breaking}.
\newblock \emph{Journal of Fluid Mechanics}, 429:\penalty0 343--380, feb 2001.
\newblock \doi{10.1017/S0022112000002846}.

\bibitem[Sutherland(2006)]{Sutherland2006}
B.~R. Sutherland.
\newblock {Weakly nonlinear internal gravity wavepackets}.
\newblock \emph{Journal of Fluid Mechanics}, 569:\penalty0 249--258, 2006.
\newblock \doi{10.1017/S0022112006003016}.

\bibitem[Tabaei and Akylas(2007)]{Tabaei2007a}
A.~Tabaei and T.~R. Akylas.
\newblock {Resonant long-short wave interactions in an unbounded rotating
  stratified fluid}.
\newblock \emph{Studies in Applied Mathematics}, 119\penalty0 (3):\penalty0
  271--296, 2007.
\newblock \doi{10.1111/j.1467-9590.2007.00389.x}.

\bibitem[Whitham(1974)]{Whitham1974a}
G.~B. Whitham.
\newblock \emph{{Linear and Nonlinear Waves}}.
\newblock John Wiley {\&} Sons, Inc., 1974.
\newblock ISBN 0471940909.
\newblock \doi{10.1002/9781118032954}.

\end{thebibliography}

\end{document}